\documentclass[twocolumn,aps,10pt,longbibliography,showkeys]{revtex4-2}
\usepackage{amsmath}
\usepackage{bm}
\usepackage{amssymb,latexsym,mathrsfs}

\usepackage{url}
\usepackage{color}

\usepackage{tikz}
\usepackage{xcolor}
\usepackage[bookmarks=false,breaklinks]{hyperref}

\definecolor{royalbluecs}{rgb}{0.2, 0.2, 1}

\hypersetup{linktoc=all, 
    colorlinks=true, linkcolor={royalbluecs},
    citecolor={royalbluecs}, urlcolor={royalbluecs}}

\definecolor{lime}{HTML}{A6CE39}
\DeclareRobustCommand{\orcidicon}{%
	\begin{tikzpicture}
	\draw[lime, fill=lime] (0,0) 
	circle [radius=0.16] 
	node[white] {{\fontfamily{qag}\selectfont \tiny ID}};
	\draw[white, fill=white] (-0.0625,0.095) 
	circle [radius=0.007];
	\end{tikzpicture}
	\hspace{-2mm}
}

\foreach \x in {A, ..., Z}{%
	\expandafter\xdef\csname orcid\x\endcsname{\noexpand\href{https://orcid.org/\csname orcidauthor\x\endcsname}{\noexpand\orcidicon}}
}

\long\def\comment#1{} 

\normalsize
 
\begin{document}

\title{Testing the Quantum of Entropy}

\author{Uwe Hohm \orcidB{}}
\email{u.hohm@tu-braunschweig.de, corresponding author.}
\affiliation{Institut für Physikalische und Theoretische Chemie, Technische Universität Braunschweig, 38106 Braunschweig, Germany}

\author{Christoph Schiller \orcidA{}}
\email{cs@motionmountain.net}
\affiliation{Motion Mountain Research, 81827 München, Germany}

\date{June 2023} 

\begin{abstract}
\noindent 
Experimental and theoretical results about entropy limits for macroscopic and single-particle systems are reviewed. It is clarified when it is possible to speak about a \textit{quantum of entropy}, given by the Boltzmann constant $k$, and about a \emph{lower entropy limit} $S \geqslant k \ln 2$.  Conceptual tensions with the third law of thermodynamics and the additivity of entropy are resolved. Black hole entropy is also surveyed. Further claims for vanishing entropy values are shown to contradict the requirement of observability, which, as possibly argued for the first time, also implies $S \geqslant k \ln 2$. The uncertainty relations involving the Boltzmann constant and the possibility of deriving thermodynamics from the existence of a quantum of entropy enable one to speak about a \emph{principle} of the entropy limit that is valid across nature.


\end{abstract}

\maketitle

\section{Introduction} 

\noindent 
In thermodynamics, the concept of the \textit{quantum of entropy} 
is rarely mentioned. 
Only a small number of authors have suggested that the Boltzmann constant $k$ plays the role of a quantum of entropy that contains and implies all of thermodynamics. 
Notable examples include Zimmermann \cite{zimo1,zimo2,zimo3,zimo4,zimo5} and Cohen-Tannoudji \cite{COHENT}.

In this article, we argue 
that the Boltzmann constant 
$k \approx 1.4 \cdot 10^{-23} \rm \,J/K$  introduced by Planck is not merely a conversion factor relating energy and temperature, but that it has a deeper meaning in nature: it fixes, with a prefactor $\ln 2$, the \textit{lower limit} to entropy.
We show this by exploring two questions. 

First, is there a `quantum of entropy' in nature? 
In particular, we test whether systems have a lower limit for entropy and whether systems can have quantized entropy values.

Secondly, can thermodynamics be deduced from the Boltzmann constant $k$
in the same way that special relativity is deduced from the speed of light $c$ and quantum theory is deduced from the quantum of action $\hbar$?
In other words, we test whether there is a \textit{principle} of the entropy limit.

In the past, different authors have arrived at different conclusions. 
We review the published  arguments in favour and against, give an overview of the results from low-temperature physics up to quantum gravity, and conclude with a structured and coherent 
summary of the quantum of entropy and its domain of application. In section 
\ref{vii}, system entropy is even shown to be bounded \textit{by definition}, using an argument that seems to be new.  In the discussion, we find it often useful to be somewhat imprecise and to call both $k$ and $k \ln 2$ the `quantum of entropy'.
We start with an issue that appears to provide a negative answer to both questions about the quantum of entropy.







\section{The third law of thermodynamics}

\noindent 
Starting in 1905, Nernst deduced a theorem about entropy that today is called the third law of thermodynamics \cite{nernst1905}. 
It has been formulated in various equivalent ways \cite{Klimenko2012}. 
Two of Nernst's formulations are:
\begin{quotation}
\noindent The \textit{entropy change} associated with a chemical or physical transition between condensed phases approaches \textit{zero} when the temperature approaches absolute zero.
\end{quotation}
\begin{quotation}
\noindent
Absolute zero temperature cannot be reached in a finite number of steps. 
\end{quotation}
In the years following its discovery, the third law was rephrased in additional ways. 
A frequent one is:
\begin{quotation}
\noindent The \textit{entropy} of a system approaches a \textit{constant value} when its temperature approaches absolute zero.
\end{quotation}
Another formulation proposed by Planck is even more popular. 
He presented it in reference \cite{Planck1954}
\footnote{%
Die Entropie eines jeden chemisch homogenen (\S 67), 
dauernd ungehemmt im inneren Gleichgewicht befindlichen Körpers von
endlicher Dichte nähert sich bei bis zum absoluten Nullpunkt abnehmender
Temperatur einem bestimmten, vom Druck, vom Aggregatzustand usw.
sowie von der speziellen chemischen Modifikation unabhängigen Wert.
Da die Entropie bisher nur bis auf eine willkürliche additive Konstante
definiert ist, können wir unbeschadet der Allgemeinheit diesen Grenzwert
gleich Null setzen. 
Translation by the authors:
The entropy of any chemically homogeneous (\S 67), 
body of finite density and in permanent uninhibited inner equilibrium approaches, when the temperature decreases to absolute zero,
a specific value that is independent of pressure, of physical state etc.,
and of the particular chemical modification.
Since the entropy was defined so far only up to an arbitrary additive constant
is defined, we can set this limit, without prejudice to generality,
equal to zero.}
and in reference
\cite{Planck1912}
\footnote{%
Dieselbe besagt, daß die Entropie
eines kondensierten (d.h. festen oder flüssigen) chemisch einheitlichen
Stoffes beim Nullpunkt der absoluten Temperatur den Wert Null besitzt [...].
Translation by the authors:
It states that the entropy of a condensed (i.e., solid or liquid) chemically homogeneous
substance has the value zero at the zero point of the absolute temperature [...].}%
, stating:
\begin{quotation}
\noindent At zero temperature, the \textit{entropy} of a chemically homogeneous body in equilibrium is \textit{zero}.
\end{quotation}
%
In this last formulation, the details are important. 
For example, in the case of glasses, which have a high configuration entropy, equilibrium is not attained, and the third law can only be formulated with the unattainability of zero temperature \cite{glasses}.
As another example, a crystalline solid is stated to have zero entropy only if it is perfect, without any impurities, dislocations, or any other crystal defects, and with all nuclear spins locked against each other. 
In addition, all formulations of the third law are valid in the \textit{thermodynamic limit,} i.e., for systems that have an infinite number of particles, infinite volume, but constant density.

Presentations, summaries and research on the third law of thermodynamics are found, for example, in references \cite{marquet,VANEKEREN199875,HONIG20071,masanes, uffink2017masanes,Hirschfelder1954,Tolman1980,Gyftopoulos2005,Sommerfeld1977,Reif1965,Klotz2008,Eastman1936}. 
In the present context, the \textit{precision} of the predicted \textit{zero} values appearing in the various formulations of the third law is central.
Given that the third law was derived using the thermodynamic limit and classical physics, several questions arise.


--\ Is the third law confirmed by experiments?

--\ Is the third law valid for small systems, and in particular, for single particles?

--\ Is the third law valid in quantum theory?

\noindent The exploration will show that for each question there are systems that do \emph{not} follow the naive third law stating that entropy can vanish at vanishing temperature.
Systems never actually have vanishing temperature 
and there indeed is a `quantum of entropy' in nature.
A first hint arose already long ago.

\section{A smallest entropy value?}
\label{sec:6op}

\noindent 
Szilard was the first researcher to suggest, in 1929 \cite{szilard}, 
that in single-particle systems, a quantum of entropy 
occurs in nature and plays an important role.  
Using simple thought experiments while exploring the details of Maxwell's demon, he deduced the value $k\,\ln 2$ for the entropy \emph{change} in the case that a free particle is forced to choose between two possible enclosed volumes of the same size. 
The numerical factor $\ln 2 \approx 0.69314...$ is due to Boltzmann's expression for entropy $S= k \ln \Omega $ in a situation where the particle chooses between two equal volumes.
In modern language, the factor $k \ln 2$ expresses that the quantum of entropy in a measurement with two microstates is described by a single bit of information. 

Szilard thus explained that there is a quantized entropy \textit{change}, i.e., a finite entropy \textit{step} in nature, whenever a \textit{single} particle changes from a situation with one possible state to a situation with two possible states, yielding an entropy change of $k \ln 2$. 
Exploring the measurement process performed by Maxwell's demon, he writes 
``größer dürfte die bei der Messung entstehende Entropiemenge freilich immer sein, nicht aber kleiner'', or ``the amount of entropy arising in the measurement may, of course, always be greater [than this fundamental amount], but not smaller''.
His description needs several clarifications.

-- Szilard does not discuss entropy values \textit{per particle} in multi-particle systems.  He leaves open whether a {smallest} or largest entropy {value} or entropy {change} {per particle} exists in nature. 

-- Szilard does not discuss the entropy of \textit{macroscopic systems.}  He {leaves open} whether a \textit{smallest} system entropy \textit{value} exists in nature. 
%
Szilard also leaves open whether a {smallest} value for the entropy \textit{change} for a macroscopic system exists in nature. 


-- Szilard discusses the case of a \textit{one-particle system} with a small number of microstates. He suggests a \textit{characteristic} value for the entropy \textit{change} for small numbers of microstates.
He does show -- when discussing the first equation in his paper -- that a smaller value of entropy does not arise.
In contrast, high numbers of microstates allow both smaller and larger values of entropy change.



Generally speaking, Szilard highlights the relation between entropy steps and the quantization of matter. 
Without particles, entropy steps would not occur, and the Boltzmann constant $k$ would not arise.

Not long ago, Szilard's thought experiment has been realized in the laboratory by Koski et al. \cite{koski2014experimental}. 
Also the quantum thermodynamics experiments based on quantum dots by 
Durrani et al. \cite{durrani2022room},
Abualnaja et al. \cite{abualnaja2022device} and 
those based on nuclear magnetic resonance by 
Vieira \textit{et al.} \cite{Vieira2023} 
confirm Szilard's results, including the value $k \ln2$.


\textit{In short,} the paper by Szilard does not make clear statements on the 
importance or the application of a possible \textit{quantum of entropy} or \textit{limit entropy} $k\,\ln 2$. It still needs to be established whether and under which conditions Szilard's value -- or a similar value such as $k$ itself -- is a useful concept for describing nature. 
This exploration can be divided into six cases.

The quantum of entropy must be compared to the observed values of (1) the entropy and (2) the entropy change \textit{per particle}, in macroscopic systems.
%
Then, the quantum of entropy must be compared to observed values of (3) the {entropy} and (4) the entropy change of a \textit{single particle}.
%
Finally, the quantum of entropy must be compared to observed values of (5) the total {entropy}   
and of (6) the entropy change for \textit{a large system}.
For each option, numerous modern experiments provide detailed observations,
and various calculations provide interesting insights. This exploration approach avoids making too general statements about a quantum of entropy too quickly 
\cite{fernandez2016holographic}.

\medskip

\section{The entropy per particle in macroscopic systems}

\noindent
In experiments with \textit{macroscopic matter} systems at low temperatures, entropy values \emph{per particle} much lower than $k\,\ln 2 \approx 0.69\,k$ have been measured. 
For instance, while lead has an entropy per atom of $7.79\,k$ at room temperature, diamond has an entropy per atom of $0.29\,k$, which is lower than the proposed quantum of entropy. 
At a temperature of 1\,K, solid silver has an entropy per atom of $8.5 \cdot 10^{-5}\, k$ \cite{leff,silverlow}. 
Similarly, in Bose-Einstein condensates, entropy values per atom have been measured to be as low as $0.001\,k$ \cite{naturek}, with a total entropy of about $1000\,k$ per million particles. 
It is planned to achieve even lower values for the entropy per particle in future microgravity experiments \cite{frye2021bose}.
Fermion condensates show similarly small values for the entropy per atom \cite{khodel2005curie}.
Also superfluid helium-II can be cited as an example of a system with an almost negligible entropy per particle \cite{Glick1969,Mongiovi1991}.
As another example, ${}^3\rm He$ has, in the region between 0.01\,K and 1\,K, an entropy of at least $k \ln 2$ per atom, due to the nuclear spins; however,  
at much lower temperatures, when the material solidifies and the spins interact, the entropy is much lower \cite{Halperin1978PropertiesOM}. 

Calculations of entropy in specific atomic systems confirm the experiments just cited. For \textit{gases,} using the results of 
Białynicki-Birula and Mycielski \cite{IBB1975},
Gadre and Bengale \cite{gadre1987rigorous} and  Angulo and Dehesa
\cite{Angulo1992} derived the entropic uncertainty relation for a single atom 
	\begin{equation}
		S_\rho+S_\gamma \geqslant 3 (1+ \ln \pi) k 
	\end{equation}
	where $S_\gamma$ and $S_\rho$ are the momentum-space and position-space information entropies, respectively.

In contrast, in  \textit{solids,} the entropy per particle can be \textit{much} lower

than $k$.
The calculations are more involved, but confirm the observations. 
Ludloff \cite{Ludloff1931a,Ludloff1931b,Ludloff1931c} used quantum statistics to deduce $S(T=0)=0$ for macroscopic bodies.
Dandoloff and Zeyher \cite{Dandoloff1981} determined that the entropy per particle approaches zero as $T \to 0$.
A similar result was obtained by De Leo et al. \cite{de2011thermodynamics}.
 
Also the entropy of \textit{photon ensembles} has been explored in detail.
Given the observation that light carries entropy, any experiment with light that shows photon behaviour can be used to deduce that individual photons carry entropy. 
However, the value of the quantum of photon entropy needs to be determined.

The entropy of the \textit{black body photon gas} has been presented by many researchers, for example in the references \cite{Sommerfeld1977,zimo2,leff2002teaching,nagata2019another}.
For example, Zimmermann explained that the entropy $S$ of a black body photon gas with $N$ photons is strictly larger than $k N$ because the entropy is not only due to the (average) number of photons $N$ but also due to their momentum distribution.
The result for the thermal photon gas is 
\begin{equation}
S=\frac{2 \pi^4}{45\, \zeta(3)} k N \approx 3.602\,  k N  \;\;,
\end{equation}
$\zeta(3)$ being the Riemann zeta function.

The situation changes in \textit{beams of light}, where all photons have the same frequency and similar directions \cite{ore1955entropy}. 
Scully \cite{scully2019laser} estimated the entropy change of a laser with one \textit{additional} photon at the threshold and found that the change can easily be as low as $ 10^{-6}\, k$.
However, as pointed out by Li et al. \cite{liliyang}, the definition of single-photon entropy is involved and not unique. 
At equilibrium, the entropy $s$ of a single monochromatic photon with frequency $\omega$ can be argued to be $s = \hbar \omega/T$. According to this relation, for visible light at room temperature, one gets $s \approx 100 k$.
At the same time, the full entropy $S$ of such a light field can be calculated to be of the order of $ S= (1+N)\ln(1+N) k - N \ln N k\approx \ln (N+1) k$.
As a result, the entropy $s$ per photon in a monochromatic light beam with large $N$ is  $s \approx k \ln N/N$, which is again much smaller than $k$.  

\textit{In short,} in macroscopic, multi-particle systems, experiments show that the concept of a quantum of entropy \emph{neither} implies a smallest value for the \emph{entropy per particle}, \textit{nor} does it imply a smallest value for the \textit{entropy change per particle.}
This result is as expected from thermodynamics: 
when the particle number increases, the entropy steps, or entropy changes, decrease without any positive lower limit.

This section thus covered the cases (1) and (2) given in section \ref{sec:6op}, with the result 
that in \textit{macroscopic} systems -- thus in the thermodynamic limit -- \textit{no} entropy 
quantum is observed, 
neither for entropy per particle nor for entropy change per particle. 
This 
result holds for both matter and radiation in the thermodynamic limit.
Therefore, we now turn to single-particle systems.

\section{Entropy and entropy change in single-particle systems}

\noindent
The central statement of this article is that a quantum of entropy exists and plays a role in single-particle systems. 
This statement must be tested in experiments for the cases of radiation and matter.

The connection between \emph{radiation} and entropy has several experimental aspects.
The results cited above imply that single photons carry entropy. 
Indeed, photon entropy has been studied in the context of laser radiation, photosynthesis, and the laser cooling of matter.
Kirwan \cite{kirwan2004intrinsic}, Van Enk and Nienhuis \cite{vanenk} \footnote{``The [produced] entropy per photon is therefore [...]  larger than k.''},
and Chen et al. \cite{chen2008recent,liu2009comment,chen2010reply}
argued in detail that a single photon carries an entropy quantum of at least $k$.

Meschke et al. observed that heat transport via a sequence of single photons confirms the quantization of heat conductance \cite{Meschke2006}.
They found that each photon carries an entropy of the order $k$.

Entropy and information are related.
The physics of information was studied already by Brillouin. 
In his influential book \cite{BRILL}, he explored the idea that the photon is a quantum of information that carries an entropy $k$.

All the mentioned publications confirm that a \textit{single photon} that is \textit{not} part of a photon field always carries a quantum of entropy of the order of $k$.

In the case of \textit{matter}, particularly for matter systems with only two possible states, numerous experiments have been performed. 
This topic became popular in 1961 when Landauer stated that the entropy required to erase one bit of information is at least $k \ln 2$ \cite{landauer1961irreversibility}. 
In the years since his publication, the limit has been confirmed for both optical and magnetic storage systems. 
Extracting entropy or erasing information from a macroscopic system requires energy. 
For a \textit{memory with one bit}, and thus two states, the required energy is defined by the entropy $k \ln2 $ and the temperature of the system.
Numerous experiments with glass beads in a double-well potential and with many other systems have confirmed this entropy value within the experimental uncertainties. 
An overview is provided in the book \cite{lent2019energy}; 
specific experiments are presented in references \cite{berut2012experimental,jun2014high,hong2016experimental}.

In addition, experiments about \textit{entropy flow} have a long history.
Many experiments during the past decades detected 
quantized thermal conductance in multiples of $\pi^2k^2T/3h$ ($T$ being the temperature) and quantized entropy conductance in multiples of $\pi^2k \nu/3$ ($\nu$ being the carrier frequency). 

A selection of experimental observations of quantized entropy flow using \textit{phonons} can be found in references \cite{schwab1,schwab2,schwab3,jezouin2013quantum,Partanen2016,cui2017quantized,mosso2017heat}. 
Quantized flow was also observed for \textit{electrons} \cite{molenkamp1992peltier,chiatti2006quantum} and  \textit{anyons} \cite{banerjee2017observed}.
The numerical value of the quantum limit of heat flow for electrons was confirmed experimentally in 2013 \cite{jezouin2013quantum}.
All these experiments confirmed that for a single quantum channel, quantum effects provide a lower limit for 
heat and entropy flow.

Calculations of the quantized entropy conductance were discussed in detail by Márkus and Gámbar \cite{ferenc} and Strunk \cite{strunk2021quantum}. 
Their analyses confirmed that for single channels, entropy transport is quantized.
The experiments and theory of entropy conductance in matter thus yield results similar to the results for entropy transport by single photons mentioned above.

Also in two-dimensional electron gases, quantized entropy per particle was first predicted and then observed \cite{PhysRevB.93.155404,grassano2018detection}.

Further investigations have explored \textit{information flow}. 
In 1983, Pendry showed that information flow is entropy flow divided by $k\,\ln2$ \cite{pendry1983quantum}. 
He also showed that in quantum channels, entropy flows, heat flows and energy flows are connected, 
a relationship that was  further clarified by Blencowe and Vitelli \cite{blencowe2000universal}.
Nowadays, entropy per photon is even used to configure optical networks \cite{Kilper2020}.


\comment{
On the other hand, in \textit{single-particle systems,} entropy \textit{is} always larger than the Boltzmann constant $k$ or than $k \ln 2$. 
All these results are consistent; entropy depends on the number of accessible quantum states. 
Every added \textit{quantum state} produces a change of entropy; in a large system, the entropy change can be extremely small, whereas it is of the order of $k$ in a single-particle system. 
}

Theoretical single-particle thermodynamics does not appear to have explored the topic of a smallest entropy of single particles.
For example, Bender et al. deduced that a heat engine for a single particle can be realized \cite{Bender2000,bender2002entropy}.
However, they deduced no statement about the existence of a quantized entropy change, neither in favour of nor against it\footnote{C.M. Bender, private communication} \footnote{D.C. Brody, private communication}. 
Also the experimental realization \cite{rossnagel2016single} of a single-atom heat engine makes no such statement.
%
Some treatments of single-particle thermodynamics explicitly disagree, such as Ali \textit{et al.} \cite{Ali2020}, who stated that the entropy of a single particle vanishes at zero temperature when it is coupled to a bath. We resolve this issue below.

\comment{
To be thought through. Is it an issue? Is it important? The next question is to clarify the appearance of slightly different values for the quantum of entropy.
The numerical values of the conductance quanta differ because ... 
not sure that correct. 
}

\textit{In short,} experiments confirm that a quantum of entropy exists and is observable in single-particle systems, in full contrast to the case of the thermodynamic limit.
In the case of single photons, single phonons, single electrons, single atoms, and single glass beads, a quantum of entropy is observed:
\begin{quotation}
\noindent$\rhd$ {Single particles carry a finite entropy that is never lower than $k \ln2 $.}
\end{quotation}
This is an important experimental finding.
\textit{An entropy limit exists because radiation and matter are made of particles.}
The statement can be falsified: to do so, it suffices to measure a system with a smaller total entropy than $k \ln 2$.
We note that the result places no limit on entropy \textit{changes} or \textit{steps}. 
These can be arbitrarily large or infinitesimally small. 
This summary settles the cases (3) and (4)
given in section \ref{sec:6op},
which asked about the entropy of single particles.
The final step is to check whether single particle entropy is also the lower limit 
for the total entropy of large systems, particularly because the third law appears to contradict this limit.



\section{Quantum theory and the third law}

\noindent 
The final cases (5) and (6) listed in section \ref{sec:6op} are the exploration of possible limits for \textit{system entropy} and for the \textit{change} of system entropy.
We begin by exploring the results of quantum theory about the third law of thermodynamics.

Already Einstein noted the necessity of considering \textit{quantum theory} to prove the third law of thermodynamics \cite{einstein1907plancksche,strehlow2005kapitulation}.
Indeed, quantum theory appears to confirm that the entropy of a condensed matter system vanishes at zero temperature, provided that its ground state is unique, and thus not degenerate \cite{Ludloff1931a,Ludloff1931b,Ludloff1931c}. 
Also Wehrl, in his influential review, 
stated that the entropy of a pure quantum state is exactly zero \cite{Wehrl1978}.
These authors concluded that $S(T=0)=0$ for degeneracy $g=1$. 
Dandoloff and Zeyher argue that a perfect crystal in its ground state has only a single microstate and therefore has vanishing entropy 
\cite{Dandoloff1981}. 
The same point was made using the modern approach of quantum thermodynamics \cite{Ali2020}.
These results were deduced in the thermodynamic limit.

The validity range of the third law has been explored by several authors. An example is the discussion by Lawson \cite{Lawson1981}. He found no experimental deviation from the third law.

Pa\~{n}oz and P\'{e}rez \cite{Panos2015} compared the experimental results on entropy $S$ to the Sackur-Tetrode equation. Within the measurement uncertainties, very good agreement was observed, provided that $S(T=0)=0$ is chosen.

The study by Loukhovitski et al. \cite{loukhovitski2022toward} confirms 
that solid nanoparticles follow the third law.

The theoretical situation is similar.
Scully calculated $S(T=0)=0$ for a Bose-Einstein condensate \cite{Scully2020}.
Ben-Naim claimed that $S(T=0)=0$ in references \cite{Ben-Naim2019,Ben-Naim2020}.
The mathematical analysis by Belgiorno confirms the Planck version of the 3rd law, $\lim_{T \to 0^+}S(T)=0$ \cite{Belgiorno2003a,Belgiorno2003b}.
According to Shastry \textit{et al.} the third law is also valid in open quantum systems \cite{Shastry2019}.
Steane presented an alternative route to obtaining $S(T=0)$, without the third law or quantum mechanics \cite{Steane2016}. He also stated that $S(T=0)=0$ is observed in many cases.

In contrast to the mentioned authors, 
some theorists argue that 
the third law with its expression $S(T=0)=0$ is a convention. This was told by Klotz \cite{Klotz2008} and Falk \cite{Falk1959}. 
Griffiths goes further. He claimed that the entropy does not vanish even for ground state non-degeneracy \cite{Griffiths1965} because the microstates near the ground state also play a role in the calculation of entropy.  Aizenman and Lieb disagreed and argued that the validity of the third law is decided completely in terms of ground-state degeneracies alone \cite{Aizenman1981}. However, they also stated that their argument is not completely tight.

\textit{In short,} experiments and many theoretical publications \textit{in the thermodynamic limit} confirm the third law of 
thermodynamics, with its zero entropy and zero entropy change (within the measurement limits) at zero temperature. 
In contrast, quantum theory does \textit{not} confirm the third law in general, 
and in particular, quantum theory does \textit{not} confirm the third law for \textit{finite} 
systems, i.e., for systems that do \textit{not} realize the thermodynamic limit.
Indeed, \textit{exact} zero entropy at zero temperature \textit{cannot be achieved} in finite systems, as argued now.

\section{The lower limit on system entropy -- and observability}
\label{vii}

\noindent
As just summarized, 
\textit{no} experiment or study has ever found a deviation from the third law for macroscopic systems. 
In other terms, in the thermodynamic limit, entropy and entropy change are effectively \textit{continuous} quantities.
%
%

\comment{
The observation of total entropy for macroscopic systems at low temperatures remains difficult. 
For macroscopic systems, no measurement that can distinguish a total entropy value $S=0$ from the value $S= k \ln 2$ has ever been published.
For example, experiments such as those by Paños and Pérez \cite{Panos2015} cannot distinguish reliably between the two cases. }


In contrast, for \textit{small} systems, all experiments -- such as the experimental tests of Szilard's experiment and of information erasure listed above -- and most calculations agree on a lower entropy limit.
When one bit of information is erased, the entropy increases by $ k \ln 2$.
Also according to Ben-Naim, the Shannon measure of information is connected to the entropy by applying a factor of $k \ln 2$ \cite{Ben-Naim2019}. Nevertheless, one can still argue that these results are not convincing, because the concept of information does not occur in quantum mechanics.

A general argument is regularly provided against a non-vanishing lower limit for the entropy of macroscopic systems. 
It is regularly stated that a quantum system in a \textit{non-degenerate} ground state 
does have vanishing total entropy: 
the expression $S=k \ln \Omega$, where $\Omega$ is the number of microstates, implies that $\Omega=1$ and thus the zero-point entropy vanishes exactly. 
However, there are at least two reasons why this popular argument is incorrect.

First, given the measurement uncertainties in the measurement of any (quasi-) continuous quantity, one \textit{cannot} prove that the quantity has a zero value.   
This is especially the case for a quantity such a entropy which is, by definition, always positive.
Quantum theory always yields non-zero measurement uncertainties, 
also for entropy and temperature.
These measurement uncertainties are related to Boltzmann's constant $k$, as shown later on.
Measurement uncertainties imply that an exactly vanishing entropy value cannot and does not exist.

Secondly, any physical system has a basic property: it is \textit{observable.}
Any observation is an interaction.
For example, observing a car implies scattering photons from it.
As another example, observing a mass can mean placing it on a scale.
Every observation and every measurement requires an interaction with the measurement apparatus.
The interaction 
implies that the system being observed has \textit{several} microstates. 
%
In particular, the basic property of observability implies that every physical system has at least \textit{two} states: it is being observed and it is not. 
In other words, 
\begin{quotation}
\noindent $\rhd$ Observability implies a smallest entropy value of $k \ln 2$ for every system.
\end{quotation}
In other terms, the case with state multiplicity $\Omega=1$ is \textit{impossible} for an \textit{observable} system.
A striking way to put this result is the following: \textit{only an unobservable system can have zero entropy.}


The argument just given resembles the well-known statement by DeWitt \cite{DEWITT1975295}. 
He stated that every system is either there or not, and that therefore, any system must have at least an entropy $k \ln 2$. To the best of our knowledge, 
the argument based on observability is not found in the literature.

Only a system that is never observed and never interacts with the environment could have vanishing entropy.
However, no such system exists, because these conditions contradict the concept of `system'. The conditions even contradict the concept of objectivity:
Unobservable or non-interacting systems are not part of the natural sciences.

The number of publications mentioning the smallest entropy value in nature is surprisingly small. 
Natori and Sano
\cite{natori1998scaling},
Ladyman et al. \cite{ladyman2007connection} or Norton \cite{norton2014brownian}, who explore the entropy of computation, prefer to state that the limit applies to entropy change. 
However, as mentioned above, this result is questionable, particularly when the number of microstates $\Omega$ is large and changes by only a small value. 
%
Observability does not seem to allow deducing a smallest value for entropy \textit{change}. 
This impossibility is an expected consequence of the third law of thermodynamics. 

It should be mentioned that while experimentally, entropy is a uniquely defined concept, in theoretical physics one can explore
Shannon entropy, von Neumann entropy,  Tsallis entropy, and Renyi entropy.
While most of these types of entropy seem to be bounded by $k \ln 2$ for systems of one particle, the present work makes this statement first of all for experimentally measured physical entropy values.

\textit{In short,} experiments and fundamental arguments confirm 
\begin{quotation}
\noindent $\rhd$   There is a lower limit for system entropy given by $S \geqslant k \ln 2$. 
\end{quotation}
The result is valid generally, for single-particle systems and for macroscopic systems.
The entropy limit also follows from the possibility to observe any physical system. The limit on system entropy is valid by definition and is independent of the substance and degrees of freedom of the system. This argument is not found in the literature so far.
In contrast, we saw above that the entropy limit does \textit{not} apply 
to the entropy \textit{per particle}, \textit{nor} to entropy \textit{steps}.
However, a further counter-argument to the quantum of entropy must still be clarified.

\section{The minimum entropy vs. the extensivity of entropy}

\noindent
A minimum entropy value can appear paradoxical because  
a minimum value seems to contradict the extensivity of entropy.
Because the minimum entropy also applies to a single atom, an everyday system, composed of many atoms, should have an entropy value given by the minimum entropy multiplied by the number of atoms. 
However, as discussed above, experiments show that this is \textit{not} the case: 
the entropy per particle can be much lower than $k \ln 2$.

This paradox was already known to Gibbs, as expalained by Jaynes \cite{jaynes1992gibbs}.
Jaynes tells that Gibbs understood that
`when two systems interact, only the entropy of the whole is meaningful. Today we would say the interaction induces correlations in their states which makes the entropy of the whole less than the sum of entropies of the parts.'

The work 
by Tsallis \cite{tsallis2006entropy} makes the same point.
In this and his other papers on the
non-extensivity of entropy, Tsallis shows that 
the extensivity of entropy requires certain conditions on the states of the subsystems: the subsystems must not be correlated.
These conditions are not fulfilled when single particles are composed to form a solid.
In other words, entropy is extensive only in the thermodynamic limit.

\textit{In short,} 
extensivity does not contradict the existence of a lowest total system entropy.
So far, the discussion of all possible lower limits 
for the entropy of single particles and for system entropy therefore confirms that 
there is a lower limit for system entropy, with the value $k \ln 2$. 
We can now explore the next question posed at the beginning.

\section{Is total system entropy quantized?}

\noindent
When Planck explored black body radiation, he discovered, introduced and named both the quantum of action~$\hbar$ and the Boltzmann constant $k$. 
Continuing our exploration, we can ask whether \emph{total system entropy} is quantized, i.e., whether its value, 
even when macroscopic, is a multiple of the quantum of entropy.

The idea of the quantization of total entropy is suggested by analogy with thermal energy.
Thermal energy can be condidered as a multiple of $kT/2$.
For total entropy however, the expression $S = k \, \ln \Omega$ implies the lack of entropy steps.
Also experimentally, entropy steps of the order of $k$ in macroscopic systems have not been detected.
Indeed, neither theoretical nor experimental claims about the issue are found in the literature.

The closest claim to the topic has been made for materials with a small amount of disorder at low temperatures.
In this case, observable entropy steps have been predicted \cite{saito2005configurational}. 
However, so far, no experiment confirmed the prediction.

The idea of the quantization of total entropy can also arise from an analogy with black holes, where total entropy is indeed 
quantized, as a result of the quantization of the area in multiples of the Planck area.
However, no such argument arises for three-dimensional systems at everyday scales in flat space. 

\textit{In short,} the total entropy of three-dimensional systems 
composed of a 
macroscopic number of particles is \textit{not} quantized in a practical sense. 
In experiments, the quantum of entropy only arises in systems with \textit{small} numbers of degrees of freedom.
However, there is one exception.

\section{Black hole horizons}

\noindent
In the domain of \textit{quantum gravity}, the entropy of gravitational 
horizons -- as they arise in black holes -- is quantized.
Many scholars have explored the quantization of black hole entropy, usually starting from the Bekenstein-Hawking entropy 
\begin{equation}
    S = k \frac{A}{4G\hbar/c^3}  \;\;. 
\end{equation}
Many authors have argued that in black holes, entropy is quantized in multiples of a smallest value, in the same way that the area of horizons $A$ is quantized in multiples of the Planck area ${G\hbar/c^3}$.
One reason for the quantization is that black holes differ from everyday thermodynamic systems because they are effectively \textit{two}-dimensional.

The value of the quantum of entropy for black hole horizons remains a matter of debate.
This value has been argued to be $k\, \ln 2$, as was done in 1975 by DeWitt \cite{DEWITT1975295}, then by Mukhanov \cite{mukhanov}, and by García-Bellido \cite{garcia1993quantum}.
As mentioned, DeWitt also argued that $k \ln 2$ is the \textit{maximum} entropy that an elementary particle can carry, because the \textit{least} information one can have about it is whether it exists or not, which is 1 bit.
Feng et al. \cite{Feng2016} came to the same conclusion by referencing Bekenstein \cite{Bekenstein1973}.

%
In contrast, Hod \cite{hod1998bohr,hod2020bekenstein} argued for an entropy quantum $k\, \ln 3$, and explained that Bekenstein also favoured this value.
Instead, Kothawala et al. \cite{Kothawala:2008in}, Skákala \cite{skakala2014quantization, skakala2014horizon}, Maggiore \cite{maggiore2008physical}, Liu et al.\cite{Liu_2009}, Ren et al \cite{ren2010entropy}, Yu and Qi \cite{Yu-Qi} and Bakshi et al. \cite{BAKSHI2017334} argued for a horizon entropy quantum of $2\pi k$.
Corishi et al. \cite{corichi2007black,corichi} proposed $2 \gamma_0 \ln 3\, k$,  where the Barbero-Immirzi parameter $\gamma_0$ is unspecified,
Sakalli et al.\cite{sakalli2012fading} and Rahman \cite{rahmana, rahmanb,rahman2020entropy} deduced more complex expressions.
Liao and Shou-Yong \cite{Liao_2004} deduced $2 \pi k/3$, and 
Jiang \cite{JQQ} and Aldrovandi and Pereira \cite{aldrovandi2008physics} deduced the value $k$.
The list is not exhaustive but gives an impression of the situation.

A different approach was used by Mirza et al. \cite{mirza}, who
showed that in black holes, the emission of entropy is limited by a value of the order of $k$ divided by the Planck time.
Given that the Planck time is the shortest time that can be measured or observed in nature, the entropy emission limit again implies the existence of a quantum of entropy of the order of the Boltzmann constant $k$.
However, no precise numerical factor has been deduced from the entropy emission limit.

The numerical prefactor in the entropy quantum in all these papers varies because, owing to the impossibility of measuring black hole entropy in experiments, a choice must be made: the number of microstates per area must be clarified.
In popular accounts, the horizon area is assumed to have one bit per Planck area ${G\hbar/c^3}$; however, this choice does not agree with the expression by Bekenstein and Hawking.
The situation simplifies drastically if one assumes an average of $e=2.718...$ microstates for each area $4G\hbar/c^3$.
In this case, the quantum of entropy for black holes is simply $k$. 
The number of microstates per horizon area can only be settled definitely with a theory of quantum gravity.
(An approach is presented in reference \cite{cs9lines}.)

In quantum gravity, also curved space far from black holes is known to contain entropy and flows of entropy \cite{jac,verlinde2011origin}. 
In contrast, infinite, flat and empty space does \textit{not} contain entropy.
However, exploring the entropy of curved space yields the same issues and results as exploring the entropy of black holes:  
the entropy of curved space cannot be measured experimentally and the calculations yield the same discussions as those for black hole horizons.

\textit{In short,} quantum gravity suggests that the full entropy of black hole horizons 
is quantized in multiples of~$ {\cal{O}}(1)k$. 
Horizons, which are essentially two-dimensional structures,  \textit{differ} from three-dimensional systems such as materials or photon gases, where macroscopic system entropy is effectively continuous and not quantized.
It must be stressed that there are no experimental data on the entropy of black holes.
There is no way to  experimentally check whether black hole entropy is quantized and, if so, what the exact value of the entropy quantum is.
Even in so-called \textit{analogue} black holes, such as acoustic black holes or superfluid ${}^4$He analogues, to our knowledge, quantized entropy has not yet been measured, even though such an effect has been predicted 
\cite{ANACLETO20161105,Herdman2017}. 
To the best of our knowledge, no discussion on the achievable measurement precision of black hole entropy has been published.

\section{Against a `quantum of entropy'}  

\noindent
Several arguments have been made against the use of the expression `quantum of entropy'.
First, this expression is used rarely.
As mentioned, the quantum of entropy is not mentioned in most textbooks on thermodynamics. 

Secondly, the concept of a quantum of entropy is confusing.
A quantum is usually considered as the \textit{smallest} possible value. 
However, in the case of entropy, a smallest value only exists for system entropy, but not for entropy steps or entropy changes, which can be extremely small.
On the other hand, energy levels in atoms behave similarly: they have quantized values, but they can be extremely close in value.


Third, system entropy is \textit{not} quantized in multiples of $k$ in any practical system -- except possibly for black holes. 
Speaking of a quantum without quantization generates uneasiness. 
On the other hand, energy in quantum systems again behaves in this way.
Like the possible energy levels in quantum systems, also the possible entropy values in thermal systems depend on the system details.
Entropy quanta are not countable in most cases -- except for the cases of quantized heat conduction.
The mentioned criticisms are also made by Blöss in his work
\cite{bloess2010entropie}. 

One might prefer the expression `lower entropy limit' to that of `quantum of entropy'. 
One could consider the expression `quantum of entrops' an example of modern \textit{hype}. 
In this article, both expressions are used.

\textit{In short,} the term `quantum of entropy' is unusual but no hard argument appears to exist against the use of the term.
If one prefers, one can use the expression `lower entropy limit' instead. 
We note that a \textit{falsification} of the quantum of entropy is straightforward: it is sufficient to measure a smaller value than $k \ln 2$ for system entropy.
However, given the tight relation between $k$ and the particle structure of matter and radiation, 
it is unlikely that this will ever happen.

In physics, an expression such as `quantum of entropy' is loaded with associations.
Its use makes only sense if it also expresses a deeper, underlying \textit{principle} of nature. 
In many domains of physics and chemistry, descriptions of natural processes using limit principles have been fruitful \cite{hohm2000there}. 
Therefore, in the remaining sections, we check in detail whether the quantum of entropy is an actual \textit{principle} of thermodynamics, 
i.e., whether the quantum of entropy can be used to derive thermodynamics.

\section{Zimmermann's principle of the entropy limit}

\noindent
Starting in the year 2000, Zimmermann explored the concept of the quantum of entropy in a series of five papers entitled `Particle Entropies and Entropy
Quanta' \cite{zimo1,zimo2,zimo3,zimo4,zimo5}.
The series builds on his earlier work \cite{zi2,zi3,zi1}.
In the first paper, Zimmermann explained that one can describe, in a many-particle system, each particle as the carrier of a part of the entropy of the system.
In the second paper, Zimmermann derived all the properties of the photon gas from the assumption of a quantum of entropy.
In the third paper, he derived the properties of the van der Waals gas from the concept of single-particle entropy.
In the last two papers, Zimmermann explored the ideal gas and the indeterminacy relation between entropy production and time.


\textit{In short,} Zimmermann argued that statistical thermodynamics -- in particular for ideal gases, real gases and photon gases -- can be deduced from the expression
\begin{equation}
	\Delta S = {\cal O}(1)\, k   \;\;
\end{equation}
for single particles. In all the cases he studied, the numerical factor is greater than 1.
Zimmermann thus argued that the Boltzmann constant goes beyond a conversion factor between temperature and energy. 
All of Zimmermann's work suggests that there is a \emph{principle of the entropy limit}.

\section{Thermodynamics from the quantum of entropy}

\noindent
Thermodynamics, as traditionally taught,
is based on a few fundamental ideas: the existence of state variables, 
the idea that heat is a form of energy, as well as the zeroth, first, second and third law 
\cite{Landsberg1956,landsberg1964mathematical,jauch1972new,leff1996thermodynamic,lieb1999physics, Gyftopoulos2005,Klotz2008,giles2016mathematical}.
Boyling makes this point particularly clear \cite{boyling1972axiomatic}.

In simple terms, statistical physics can be seen as based on the principle of least action, 
on quantum theory, and on the properties of entropy. 
The point is clearly made by Landau \& Lifshitz \cite{Landau1958} and by Kubo \cite{Kubo1965}.
Simply stated, the principle of least action implies energy conservation.
The quantum of action implies the particle structure of matter and radiation, and thus implies, together with their dynamics, the existence of temperature and other state variables.
Thus, the zeroth and first laws of thermodynamics are consequences of the principle of least action and of the quantum of action.

The existence of a quantum of entropy reproduces the equivalence of heat and energy. 
This equivalence is part of the first law \cite{mayerjoule}; it is also part of the zeroth law, i.e., of the existence of temperature.

The second and third laws of thermodynamics concern the state variable entropy directly. 
The concept of entropy is best defined and thought of as the spreading of energy
\cite{leff1996thermodynamic} or as the mixing of states \cite{seitz2022mixed}. 
%
%
The quantum of entropy includes the definition of entropy.
Simultaneously, the quantum of entropy includes the particle structure of matter and radiation. 
Using the arguments summarized in references  \cite{leff1996thermodynamic,seitz2022mixed}, the quantum of entropy implies the second law.

The third law of thermodynamics states and implies that at low temperatures, most degrees of freedom of a condensed matter system are frozen. 
Thus, the third law follows from quantum theory \cite{Ludloff1931a,Ludloff1931b,Ludloff1931c}. 
The quantum of entropy only plays an indirect role in the third law, defining the measurement unit of entropy. 

\textit{In short,} even though the topic is not treated exhaustively here, it appears that the quantum of 
entropy $k$ is at the basis of all four laws of thermodynamics.
The entropy limit is a fundamental \textit{principle} of thermodynamics. 
When the state variables, least action, and the quantum of action are included at the foundations, all of thermodynamics is covered \cite{cs9lines}.
There is a further reason to speak about the \textit{principle} of the entropy limit.

\section{Indeterminacy relations}

\noindent
Statistical physics is closely related to quantum theory. 
The relation became clear already in the early twentieth century
when the first indeterminacy relations for thermodynamic quantities were deduced. 
For example, Bohr showed that temperature $T$ and energy $U$ obey 
\begin{equation}
    \Delta (1/T) \; \Delta U \geqslant {k}/{2}.
\end{equation}
This indeterminacy relation was discussed in detail by {Heisenberg} and other scholars \cite{Uffink1999,Shalyt-Margolin:2003qig,Hasegawa:2022czt}.

In 1992, de Sabbata and Sivaram \cite{Sabbata1992} deduced the indeterminacy relation
\begin{equation}
    \Delta T \; \Delta t \geqslant {\hbar}/{k} \;\;. 
\end{equation}
This relation was tested and found to agree with experiments \cite{gillies2004experimental,gillies2005experimental}. 
%
In 2004, Kovtun, Son and Starinets showed that the ratio between shear viscosity $\eta$ and entropy volume density $s$ 
follow \cite{kss,hohm2019conjecture}
\begin{equation}
    \frac{\eta}{s} \geqslant {\hbar}/{4 \pi k} \;\;.  
\end{equation}
%
In 2011, Zimmermann showed \cite{zimmermann11} that in quantum thermodynamics, entropy production $P$ and time $t$
obey
\begin{equation}
    \Delta P \; \Delta t \geqslant {k}/{2}.
\end{equation}
Many additional indeterminacy relations exist; a comprehensive list is given by Hohm 
\cite{hohm2019conjecture}.
All these indeterminacy relations suggest that entropy resembles action, 
with a multiple $\mathcal{O}(1) \, k$ of the Boltzmann constant 
playing a role similar to that of $\hbar$.
Indeed, Parker et al. \cite{parker2021entropic,parker2022} use similar versions of the entropic uncertainty relations and the quantum of entropy to calculate the configurations of alpha particles and of cosmological systems.

Maslov \cite{maslov2002quantization} has extended the analogy between quantum theory and thermodynamics 
by defining quantum operators for internal energy, for free energy and for entropy. 
In analogy to quantum theory, the measured values of these quantities are the eigenvalues of these operators.

\textit{In short,} the quantum of entropy 
plays a similar role in thermodynamics as the quantum of action does in quantum theory.
In both cases, the minimum value arises also in indeterminacy relations.
This property again underlines that $k$ is not only a conversion factor 
but that it has a fundamental significance in thermodynamics.

\section{Similarities and differences between action and entropy}

\noindent
The \emph{similarities} between action and entropy are striking.
In nature, there exists a quantum of action $\hbar$ and a quantum of entropy $k$. 
Quantum theory, including aspects such as indeterminacy relations and entanglement, is based on the quantum of action. 
All the effects of quantum theory disappear if the quantum of action vanishes.
Thermodynamics, including the second law, is based on the quantum of entropy.
All the effects of thermodynamics disappear if the quantum of entropy vanishes.

Deducing thermodynamics from the Boltzmann constant $k$ resembles the procedure used for the new SI unit definitions by the BIPM since 2019. 
In the SI, the Boltzmann constant $k$ is used to define the international measurement unit of temperature, the kelvin K, and through it, the unit of entropy, J/K. 
The quantum of action $\hbar$ is used to define the unit of mass, the kilogram, and through it, the unit of energy, the joule J.
Without a quantum of entropy $k$, a measurement unit of entropy would not exist, 
and it would be impossible to measure temperature or entropy values.
Likewise, without a quantum of action $\hbar$, a measurement unit of mass would not exist,  
and it would be impossible to measure mass or energy values.
For the same reason, time-variations of $k$ and $\hbar$ cannot be observed, as pointed out by Duff \cite{duff2002comment},
because both quantities are central to the definitions of measurement units. 

Both action and entropy are extensive quantities.
Furthermore, both the quantum of action and the quantum of entropy are related to the discrete structure of physical systems.
Both the quantum of action and the quantum of entropy distinguish classical physics from quantum physics. 
There is a well-known continuum limit of thermodynamics -- generally consistent with the thermodynamic limit -- in which $k \rightarrow 0$
\cite{compagner1989thermodynamics}.
It leads to classical thermodynamics,
but similar to the limit $\hbar \rightarrow 0$, the limit $k \rightarrow 0$ prevents the calculation of any specific material property. 
All material properties are due to the quantum of action and the quantum of entropy. 
The quantum of action and the quantum of entropy make similar statements: 
if either quantum did not exist, particles, thermal effects and quantum effects would not exist.

The \emph{differences} between action and entropy are also important \cite{acosta2012holographic}.
Action and entropy resemble each other because they are both related to the microscopic world, but differ in their relation to \emph{change}. 
Action describes change occurring in nature as a product of energy and time; a large amount of action implies a large amount of change. 
Action is the \textit{measure} of change.
And despite the existence of a quantum of action, nature minimizes action in any process in an isolated system.
In contrast, entropy describes the spreading of energy, in particular the spreading from macroscopic to microscopic change. 
A large amount of entropy implies a large amount of spreading. 
Entropy is the \textit{cost} of change. 
Despite the existence of a quantum of entropy, nature maximizes entropy in processes in any isolated system.

The quantum of action also implies that action and also action {change} are quantized, as observed \cite{sommerfeld1911plancksche,
planck1908dynamik,sackur1913universelle,de1948max,balibar1984quantique,
hushwater1998path,
hushwater1998quantum,
sergeenko2002quantization,
pietschmann2011quantenmechanik,
curtis2004use,bucher2008rise,curtis201121st,
bartelmann2018theoretische,zagoskin2015quantum,
capellmann2017development,boughn2019wherefore,khrennikov2021devil,sergeenko2022general}. 
In contrast, this is \emph{not} the case for entropy. 
Also Parker et al. explore the similarities and differences \cite{parker2021entropic,parker2022}.


\textit{In short,} 
the quantum of action $\hbar$ \textit{implies} particles and describes their motion; 
the quantum of entropy $k$ \textit{results} from particles and describes their statistics. 
To include black holes, it can be said: 
\begin{quotation}
\noindent $\rhd$ The Boltzmann constant $k$ expresses that everything that moves is made of discrete constituents.
\end{quotation}
In other words, the quantum of entropy is a fundamental property of nature. 
The quantum of entropy is a limit of nature like the speed limit and the action limit \cite{cs9lines}.

\section{Conclusion: a consistent presentation of the quantum of entropy}

\noindent 
The present study explored several questions regarding entropy and quantization. 
In accordance with the third law of thermodynamics, it was shown that 
there is \textit{no} smallest entropy value \emph{per particle} and, likewise, \textit{no} smallest entropy 
\emph{change} -- particularly in systems consisting of a large number of particles. 
However, there \textit{is} a smallest total entropy value, a quantum of entropy, 
that is based on the Boltzmann constant $k$: 
\begin{quotation}
     \noindent $\rhd$ The lower entropy limit $S\geqslant k \ln 2$ holds for every physical system. 
\end{quotation}
\noindent
The statement agrees with all experiments and is falsifiable.
It was shown that the lower limit on entropy is only in \textit{apparent} contrast to the usual 
formulation of the third law of thermodynamics 
or to the extensivity of entropy.
In all systems with \textit{many} particles that were explored so far at low temperatures,
the total entropy value $k \ln 2$ and the total entropy value zero are indistinguishable experimentally. 
In all systems consisting of \textit{one} particle, the entropy limit is confirmed.
In contrast to the quantum of action, the quantum of entropy arises experimentally \emph{only} in physical systems with one particle.

It was found that for materials and in radiation fields, experiments and theory confirm 
that total entropy values are \emph{not} integer multiples of the Boltzmann constant $k$. 
In contrast to everyday life, in black holes, entropy \textit{is} expected to be quantized in integer multiples of the Boltzmann constant $k$.
The reason is that the underlying constituents of black hole horizons are discrete and that, in the case of black hole horizons, 
large numbers of these constituents do not result in smaller entropy steps. 
In simple terms, the \textit{quantum} of entropy holds generally, whereas countable \textit{quanta} of entropy only arise in black holes.

Finally, it was argued that the laws of thermodynamics
can be deduced from the state variables and the quantum of entropy.
The quantum of entropy also explains the indeterminacy relations between thermodynamic variables.
Thus, the Boltzmann constant $k$ is more than a simple conversion factor: 
it is a fundamental property of nature expressing that everything that moves is composed of discrete components.

In conclusion, in the same way that 
the speed limit $c$ is a principle of special relativity and 
the quantum of action $\hbar$ is a principle of quantum theory, also the entropy limit $k \ln 2$ is a principle of thermodynamics.

\section*{Acknowledgments and declarations}  
\label{sec:ack}

\noindent 
The authors thank J. Norton, A. Kirwan, E. Pérez, S. Hod, C. Sivaram, A. Kenath, C. Bender, D. Brody, L. Deng, C. Blöss, J.F. Tang,  V.P. Maslov, A. Sharipov, S. Meyer and the anonymous referees for fruitful discussions.
The authors declare no conflicts of interest and no competing interests.
No additional data are available for this manuscript.

\bibliographystyle{MSP}
\bibliography{minentr-arxiv}

\newpage
\section*{Graphic summary -- Table of contents} 

\bigskip
\bigskip
\bigskip

\noindent 
{\Huge \hspace*{2.2cm} $ S \geqslant k \ln 2 $} 

\bigskip
\bigskip
\bigskip

\noindent 
This inequality for the entropy $S$ and the Boltzmann constant $k$ is shown to be valid in general. 
The entropy limit is valid despite the third law of thermodynamics and follows from the observability of physical systems. 

The entropy limit is a \textit{principle} of nature that characterizes thermodynamics in the same way that the speed limit $v\leqslant c$ characterizes special relativity and the action limit $W \geqslant \hbar $ characterizes quantum theory.

\ 
\vspace*{8cm}

\end{document}